\documentclass[twocolumn,pra]{revtex4}

\begin{document}

\title {Time-dependent density-functional theory beyond the adiabatic approximation:
insights from a two-electron model system
}
\author {C. A. Ullrich}
\affiliation {Department of Physics and Astronomy, University of Missouri, Columbia, Missouri 65211}

\date{\today}

\begin{abstract}
Most applications of time-dependent density-functional theory (TDDFT) use
the adiabatic local-density approximation (ALDA) for the dynamical exchange-correlation potential
$V_{\rm xc}({\bf r},t)$. An exact (i.e., nonadiabatic) extension of the ground-state LDA into the dynamical regime
leads to a $V_{\rm xc}({\bf r},t)$ with a memory, which causes the
electron dynamics to become dissipative. To illustrate and explain this nonadiabatic behavior, this paper studies
the dynamics of two interacting electrons on a two-dimensional quantum strip of finite size,
comparing TDDFT within and beyond the ALDA with numerical solutions of the two-electron time-dependent
Schr\"odinger equation. It is shown explicitly how dissipation arises through multiple particle-hole excitations,
and how the nonadiabatic extension of the ALDA fails for finite systems, but becomes correct in the thermodynamic limit.
\end{abstract}

\maketitle

\section{Introduction} \label{sec:intro}

The essential ingredient of time-dependent density-functional theory (TDDFT) \cite{rungegross,tddft}, the
exchange-cor\-re\-la\-tion (xc) potential $V_{\rm xc}({\bf r},t)$, is frequently obtained using
the adiabatic local-density approximation (ALDA):
\begin{equation} \label{alda}
V_{\rm xc}^{\rm ALDA}({\bf r},t) = \left.
\frac{d [\bar{n} e_{\rm xc}(\bar{n})]}{d\bar{n}}\right|_{\bar{n} = n({\bf r},t)} ,
\end{equation}
where $e_{\rm xc}(\bar{n})$ is the xc energy per particle of a homogeneous electron gas of density $\bar{n}$.
The adiabatic approximation means that all functional dependence of $V_{\rm xc}({\bf r},t)$ on prior time-dependent
densities $n({\bf r}',t')$, $t'<t$, is ignored. Neglecting the retardation implies frequency-independent
and real xc kernels in linear response. This approach has been widely used in quantum chemistry for calculating
molecular excitation energies \cite{furche}.

The adiabatic approximation is known to work well for excitation processes in many-body systems that have a direct counterpart
in the Kohn-Sham system, such as atomic and molecular single-particle excitations. On the other hand, for more
complicated processes such as double or charge-transfer excitations the ALDA can fail dramatically
\cite{Maitra1,Maitra2}. A recent study \cite{UllrichTokatly} has shown that the ALDA can completely break
down in dynamical processes where the electronic density rapidly undergoes large deformations.

Several recent papers have addressed the question how the LDA for ground-state calculations
should be properly extended into the dynamical regime \cite{UllrichTokatly,VK,Bunner,VUC,UllrichVignale,Kurzweil,Tokatly}.
Vignale and Kohn \cite{VK} showed that a
nonadiabatic {\em local} approximation for exchange and correlation requires the time-dependent
{\em current} $j({\bf r},t)$ as basic variable (C-TDDFT). This formalism was later recast in the
language of hydrodynamics, where xc effects beyond the ALDA appear as viscoelastic stresses in the electron
liquid \cite{VUC,UllrichVignale}. An alternative nonadiabatic theory formulates
TDDFT from the point of view of an observer in a co-moving Lagrangian reference frame (L-TDDFT)
\cite{Tokatly}. In Ref. \cite{UllrichTokatly}, the technical details of C-TDDFT and L-TDDFT are critically examined
and compared.

To date, most applications of TDDFT beyond the adiabatic approximation take place in the frequency-dependent
linear-response regime. A major success of C-TDDFT was the work by van Faassen {\em et al.} \cite{Faassen}
who calculated static axial polarizabilities in molecular chains, with much improvement over the ALDA.

The C-TDDFT formalism has recently been applied to describe linear and nonlinear charge-density oscillations
in quantum wells by solving the time-dependent Kohn-Sham (TDKS) equation \cite{Wijewardane}.
It was shown that the retardation caused by the memory of the xc potential has the striking
consequence of introducing decoherence and energy relaxation, i.e., the oscillating density
experiences a damping.
The mechanism causing this behavior has been discussed by D'Agosta and Vignale \cite{DAgosta}.
Technically, dissipation arises in C-TDDFT from a velocity-dependent xc (vector) potential which breaks
the time-reversal invariance of the TDKS Hamiltonian. As a result, a system tends to
relax from a nonequilibrium initial state to an equilibrium final state with higher entropy.
But where does the dissipated energy go?

Because the system is closed and isolated (there is no coupling to a thermal bath),
the {\em total} energy should be conserved. According to Ref. \onlinecite{DAgosta},
dissipation in C-TDDFT has to be understood in the sense that
energy is redistributed between two subsystems with different sets of electronic degrees of freedom, coupled
by Coulomb interactions. In the quantum well examples of Refs. \onlinecite{Wijewardane} and \onlinecite{DAgosta},
the transfer of energy occurs from a collective motion along the confinement
direction into low-lying lateral excitations of the two-dimensional electron gas in the quantum well plane.
However, in C-TDDFT this transfer process is never directly observed, since the TDKS
equations are solved only for the electron dynamics perpendicular to the quantum well plane.

The purpose of the present paper is to give an explicit, pedagogical illustration
of the road towards dissipation in collective electronic
motion. We will consider a two-electron model system that is simple enough so that its dynamics can be treated numerically
exactly via solution of the full time-dependent Schr\"odinger equation, and compare it with TDDFT
within and beyond the ALDA. In particular, we will focus on charge-density oscillations
along one direction of the system, and how the exact calculations show that the amplitude of these oscillations changes
over time. This amplitude modulation comes from a superposition of transitions between
the ground state and singly excited states and between singly and doubly excited states,
including a coupling to the transverse degrees of freedom due to Coulomb interactions.

In ALDA, all effects involving multiple excitations are completely absent; in C-TDDFT, multiple excitations are
implicitly included, but for finite systems their contribution is
strongly exaggerated, producing an unphysical damping. Based on the insights of our simple two-electron system,
we will discuss how the dissipative behavior emerges in the thermodynamic (large-system) limit, and to what extent it is
then correctly described by C-TDDFT.

In section \ref{sec:model} we give the technical details of our two-electron model system and describe how the
full Schr\"odinger equation and the TDKS equations with and without memory are solved. Section \ref{sec:results}
gives our results and discusses the physical process of dissipation of collective charge-density oscillations.
Conclusions are given in section \ref{sec:conclusions}.

\section{Model system} \label{sec:model}

Consider two electrons on a two-dimensional (2D) quantum strip of length $L$ and width $\Delta$, positioned in the $x-z$ plane.
In the following, we will be mostly interested in situations where $L>>\Delta$.
The system has hard-wall boundary conditions at two ends of the strip, at $z=0$ and $z=\Delta$, and
periodic boundary conditions along the $x$-direction. In other words, the electrons are living on a strictly
2D surface whose topology is equivalent to that of a cylindrical tube of length $\Delta$ and circumference $L$.

\begin{figure}
\unitlength1cm
\begin{picture}(5.0,7.5)
\put(-8.75,-15.45){\makebox(5.0,7.5){\includegraphics{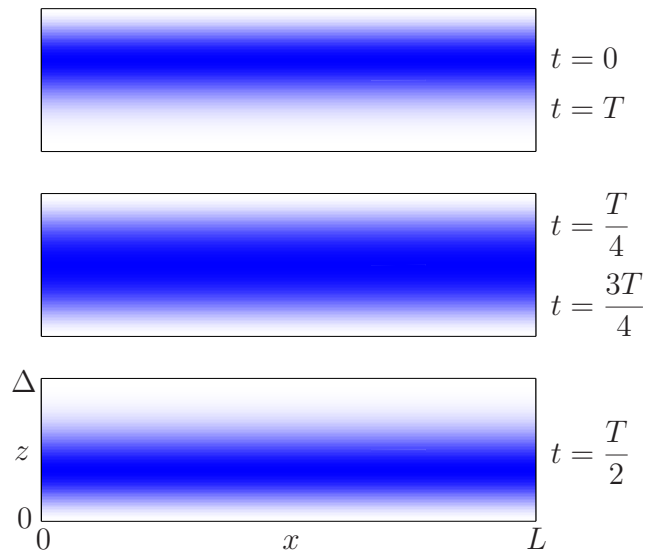}}}
\end{picture}
\caption{\label{figure1} (Color online) Schematic illustration of one cycle of a charge-density oscillation of a two-electron system on a 2D
quantum strip of width $\Delta$ and length $L$.
Darker areas represent regions of charge accumulation. Snapshots are shown at times as indicated, where $T$ is the duration of
one cycle. The model assumes periodic boundaries along $x$ and
hard-wall boundaries at $z=0$ and $z=\Delta$.
}
\end{figure}

On this 2D quantum strip we first calculate the electronic ground state in the presence of
a linear external potential which depends only on $z$:
\begin{equation} \label{external}
V(z) = Fz \:,
\end{equation}
where $F$ is a constant field strength.
At the initial time $t=0$, this external potential is suddenly switched off, which triggers a charge-density
oscillation along $z$ (see Fig. \ref{figure1}). The electronic density thus remains uniform along the $x$-direction for all times.
The goal is to follow the time evolution of the system for many cycles of the charge-density oscillations, comparing the
exact numerical solution of the two-electron Schr\"odinger equation with TDDFT solutions within and beyond ALDA. Atomic (Hartree)
units are used throughout.

\subsection{Two-electron Schr\"odinger equation}
\subsubsection{Ground state}
The static two-electron Schr\"odinger equation for our problem reads
\begin{eqnarray} \label{Schr2}
0 &=&
\left[ -\frac{\nabla_1^2}{2} - \frac{\nabla_2^2}{2} + V(z_1) + V(z_2)+ \frac{1}{|{\bf r}_1 - {\bf r}_2|} - E_j \right]
\nonumber\\
&& \times \Phi_j({\bf r}_1 s_1,{\bf r}_2 s_2) \:,
\end{eqnarray}
where ${\bf r}_{1,2} = (x_{1,2},z_{1,2})$, and $s_{1,2}$ denotes the spin. We expand the two-electron eigenstates $\Phi_j$
in a basis of Slater determinants:
\begin{equation} \label{SlaterBasis}
\Phi_j({\bf r}_1 s_1,{\bf r}_2 s_2) = \sum_{\nu_1\nu_2 \atop \kappa_1\kappa_2} C^j_{\nu_1 \nu_2 \kappa_1 \kappa_2}
\Psi_{\nu_1 \nu_2 \kappa_1 \kappa_2}({\bf r}_1 s_1,{\bf r}_2 s_2) \:,
\end{equation}
where
\begin{eqnarray}
\Psi_{\nu_1 \nu_2 \kappa_1 \kappa_2}
&=& \frac{1}{\sqrt{2}} \Big[ \psi_{n_1k_1}(x_1,z_1) \xi_1(s_1) \psi_{n_2k_2}(x_2,z_2) \xi_2(s_2) \nonumber\\
&-& \psi_{n_2k_2}(x_1,z_1) \xi_2(s_1) \psi_{n_1k_1}(x_2,z_2) \xi_1(s_2)\Big].
\end{eqnarray}
Here, $\xi$ are single-particle spinors, and for the spatial
part we choose the non-interacting single-particle wave functions for constant external potential:
\begin{equation} \label{sp1}
\psi_{nk}(x,z) = \sqrt{ \frac{2}{L\Delta}} \: e^{ikx} \sin nz
\end{equation}
with quantum numbers
\begin{eqnarray} \label{knum}
k=\frac{2\pi \kappa}{L}\:, && \kappa = 0, \pm 1, \pm 2, ...\\[2mm]
n = \frac{\pi \nu}{\Delta} \:, && \nu = 1,2,3,... \label{nnum}
\end{eqnarray}
In other words, we satisfy the given boundary conditions (see Fig. \ref{figure1}) by choosing plane wave basis states
along the strip and standing waves across the strip.

Inserting the basis expansion (\ref{SlaterBasis}) into the Schr\"odinger equation (\ref{Schr2}) yields the
following equation for the expansion coefficients:
\begin{eqnarray}\label{MatrEq}
\sum_{\nu_1\nu_2 \atop \kappa_1\kappa_2} \bigg[
(T_{\nu_1 \kappa_1} + T_{\nu_2 \kappa_2}) \delta_{\nu_1\mu_1}\delta_{\nu_2\mu_2} \delta_{\kappa_1\rho_1} \delta_{\kappa_2\rho_2}
\nonumber\\
+ (V_{\mu_1 \nu_1} \delta_{\nu_2\mu_2}
+ V_{\mu_2 \nu_2} \delta_{\nu_1\mu_1}) \delta_{\kappa_1\rho_1} \delta_{\kappa_2\rho_2} \nonumber\\
+ W_{\nu_1 \nu_2 \kappa_1 \kappa_2}^{\mu_1 \mu_2 \rho_1 \rho_2}\bigg]
C^j_{\nu_1 \nu_2 \kappa_1 \kappa_2} = E_j C^j_{\mu_1 \mu_2 \rho_1 \rho_2} \:.
\end{eqnarray}
Here, the kinetic energy and external potential matrix elements are given by
\begin{equation}
T_{\nu\kappa} = \frac{n^2 + k^2}{2}
\end{equation}
and
\begin{equation} \label{potmat}
V_{\mu\nu} = \frac{2}{\Delta} \int_0^\Delta dz \: \sin mz \sin nz \: V(z) \:.
\end{equation}
The matrix elements of the Coulomb interaction are
\begin{eqnarray}
W_{\nu_1 \nu_2 \kappa_1 \kappa_2}^{\mu_1 \mu_2 \rho_1 \rho_2}&=& \frac{4}{\Delta^2 L}
\int_0^\Delta dz_1\int_0^\Delta dz_2\,\sin m_1z_1 \, \sin m_2z_2 \nonumber\\
&\times & \sin n_1z_1 \, \sin n_2z_2 \,  \delta_{k_1+k_2,q_1+q_2} I_{k_2 - q_2}(z_1,z_2)  \nonumber\\
\end{eqnarray}
where
\begin{eqnarray}
I_{k - q}(z_1,z_2) &=& \int_{-\infty}^\infty dx
\; \frac{\cos[(k - q)x]}{\sqrt{ x^2 + (z_1 - z_2)^2}} \nonumber\\
&=& 2K_0[ |k-q| \, |z_1-z_2|] \,, \quad k\ne q \\
&=& -2 \log|z_1-z_2| \,, \quad k=q \:.
\end{eqnarray}
Here, $K_0$ is a complete Bessel function of the second kind in standard notation, and in the case $k=q$ an additional
divergent term is cancelled by the positive background.

Solving equation (\ref{MatrEq}) numerically one finds that a relatively small basis size including states with
with no more than $\kappa = \pm 10$ and $\nu = 10$ [Eqs. (\ref{knum}), (\ref{nnum})] is sufficient.
The computational task is therefore quite manageable.

Furthermore, it turns out that, due to symmetry and momentum conservation, only those two-electron
basis states $\Psi_{\nu_1 \nu_2 \kappa_1 \kappa_2}$ contribute which have zero net current along the strip, i.e.,
only states with $\kappa_1 = -\kappa_2$ are needed. This corresponds to two-electron states where one
electron travels to the right and the other to the left.

\subsubsection{Time evolution} \label{timeexact}

Once equation (\ref{Schr2}) has been diagonalized, the next step is to determine the time evolution of the ground state
$\Phi_1({\bf r}_1 s_1,{\bf r}_2 s_2,t)$ after the linear external potential has been switched off. Rather than explicitly solving
the time-dependent two-electron Schr\"odinger equation,
this is most easily done by expanding $\Phi_1$ in the complete set of field-free eigenstates, defined as follows:
\begin{equation} \label{Schr2f}
\left[ -\frac{\nabla_1^2}{2} - \frac{\nabla_2^2}{2} + \frac{1}{|{\bf r}_1 - {\bf r}_2|} - E_j^f \right]
\Phi_j^f=0 \:,
\end{equation}
\begin{equation} \label{SlaterBasisf}
\Phi_j^f = \sum_{\nu_1\nu_2 \atop \kappa_1\kappa_2} C^{j,f}_{\nu_1 \nu_2 \kappa_1 \kappa_2}
\Psi_{\nu_1 \nu_2 \kappa_1 \kappa_2} \:.
\end{equation}
Thus,
\begin{equation} \label{FieldFreeExp}
\Phi_1(t) = \sum_j A_j(t) \Phi_j^f \:,
\end{equation}
where
\begin{equation} \label{AJ}
A_j(t) = \exp[-i E_j^f t] \sum_{\nu_1\nu_2 \atop \kappa_1\kappa_2}C^{j,f}_{\nu_1 \nu_2 \kappa_1 \kappa_2}
C^{1}_{\nu_1 \nu_2 \kappa_1 \kappa_2} \:.
\end{equation}
From this, we obtain the time-dependent density as follows:
\begin{eqnarray} \label{TDdensity}
n(z,t) &=& \sum_{s_1 s_2} \int d^2 r_2 |\Phi_1({\bf r}_1 s_1,{\bf r}_2 s_2,t)|^2 \nonumber\\
&=&
\frac{2}{L\Delta}\sum_{\nu_1\nu_2 \atop \kappa_1\kappa_2} Q_{\mu_1 \mu_2 \nu_1 \nu_2}(t)
[ \sin m_1 z \: \sin n_1 z \:\delta_{\mu_2,\nu_2}\nonumber\\
&& {} +\sin m_2 z \: \sin n_2 z \:\delta_{\mu_1,\nu_1}] \:,
\end{eqnarray}
where
\begin{equation}
Q_{\mu_1 \mu_2 \nu_1 \nu_2}(t) =
\sum_{ij} A^*_i(t) A_j(t)\sum_{\kappa_1 \kappa_2}C^{i,f}_{\nu_1 \nu_2 \kappa_1 \kappa_2}
C^{j,f}_{\nu_1 \nu_2 \kappa_1 \kappa_2} \:.
\end{equation}
Finally, the time-dependent dipole moment is
\begin{equation} \label{TDdipole}
d(t) = \int_0^\Delta dz \: z n(z,t) \:.
\end{equation}

\subsection{TDDFT}
\subsubsection{Ground state}
The two-electron problem described above can be solved, in principle exactly, using the TDKS
formalism. We begin with the static Kohn-Sham (KS) equation:
\begin{equation}
\left[ -\frac{ \nabla^2}{2} + V(z) + V_{\rm H}(z) + V_{\rm xc}(z) - E_n\right] \phi_n(x,z) = 0 \:.
\end{equation}
This equation separates in $x$ and $z$, and we make the ansatz
\begin{eqnarray}
\phi_{n}(x,z) &=& \frac{1}{\sqrt{L}} \:e^{ikx} \varphi_j(z)\:,\\
E_{n} &=& \frac{\hbar^2 k^2}{2m} + \varepsilon_j \:,
\end{eqnarray}
where the index  $k$ is given by Eq. (\ref{knum}).
The ground-state solution has $k=0$, and
we end up having to solve a one-dimensional equation for $\varphi_j(z)$ and $\varepsilon_j$:
\begin{equation} \label{Seq}
\left[ -\frac{1}{2} \frac{d^2}{dz^2} + V(z) + V_{\rm H}(z) + V_{\rm xc}(z) - \varepsilon_j\right] \varphi_j(z) =  0 \:.
\end{equation}
To solve the single-particle KS equation, we expand in a standing-wave basis as follows:
\begin{equation} \label{KSbasis}
\varphi_j(z) = \sqrt{ \frac{2}{\Delta}} \sum_{\nu=1}^N C_\nu^j \sin nz \:,
\end{equation}
where
\begin{equation}
\sum_\nu \left[ \frac{n^2}{2} \: \delta_{\mu\nu} + V_{\mu\nu} + V^{\rm H}_{\mu\nu}
+ V^{\rm xc}_{\mu\nu} \right] C_\nu^j = \varepsilon_j C_\mu^j \:.
\end{equation}
From this, the ground-state density follows as
\begin{equation}
n(z) = \frac{4}{L\Delta} \sum_{\mu\nu} C_\mu^{1*} C_\nu^1 \sin mz \sin nz \:.
\end{equation}
The matrix elements for the external, Hartree and xc potential are calculated from Eq. (\ref{potmat}).
The Hartree potential is given by
\begin{equation}
V_{\rm H}(z) = -2 \int_0^\Delta dz' n(z')\log|z-z'|
\end{equation}
plus a diverging constant which is cancelled by the positive background. For the xc potential we use the
LDA within the parametrization of the 2D electron gas of Tanatar and Ceperley \cite{Tanatar}. For the (spin-unpolarized)
systems under consideration, the more modern parametrization by Attaccalite {\em et al.} \cite{Attaccalite} gives
almost identical results.

\begin{figure}
\unitlength1cm
\begin{picture}(5.0,4.9)
\put(-6.2,-10.9){\makebox(5.0,4.9){\includegraphics{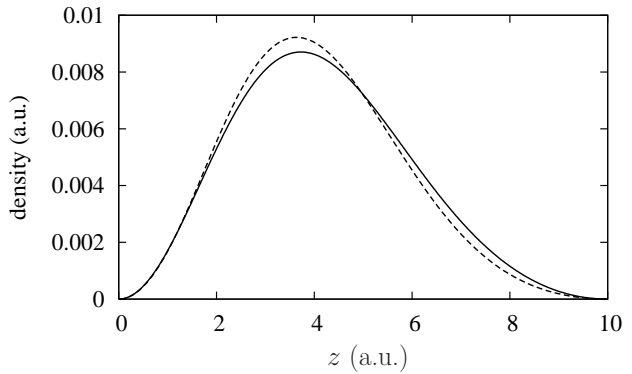}}}
\end{picture}
\caption{\label{figure2} Two-electron ground-state density $n(z)$ on a
quantum strip of width $\Delta=10$ and length $L=50$,  with field strength $F=0.02$. Full line: exact solution.
Dashed line: LDA.
}
\end{figure}

Figure \ref{figure2} shows the two-electron ground-state density $n(z)$ on a 2D quantum strip
of width $\Delta=10$ and length $L=50$, in the presence of a linear external potential (\ref{external})
with field strength $F=0.02$. For these system parameters, the 2D Wigner-Seitz radius $r_s = (\pi n)^{-1/2}$
has a value of $r_s=6$ at the maximum of the density distribution.
The agreement between the exact and the LDA density is reasonably good, and in fact becomes better for
smaller quantum strips where the density is higher. In general, the LDA system is found to be a little more
polarizable than the exact system.


\subsubsection{Time evolution} \label{timeevol}
As for the two-electron case, we set the charge-density oscillations in motion by suddenly switching off the external potential
at the initial time $t=0$. The task is to solve the TDKS equation
\begin{equation} \label{TDKS}
\left[ -\frac{1}{2} \frac{d^2}{dz^2} + V_{\rm H}(z,t) + V_{\rm xc}(z,t) - i \frac{\partial}{\partial t} \right] \varphi(z,t) =  0 \:,
\end{equation}
with initial condition $\varphi(z,0) = \varphi_1(z)$. The time-dependent KS orbital $\varphi(z,t)$ is expanded similar
to Eq. (\ref{KSbasis}), and the time-dependent expansion coefficients $C_\mu(t)$ are numerically determined using
the Crank-Nicholson algorithm plus predictor-corrector scheme \cite{tddft}.

In the following, we will consider $V_{\rm xc}(z,t)$ within and beyond the ALDA.
The C-TDDFT expression for a time-dependent xc potential with memory is called ALDA+M, and
written as follows \cite{UllrichTokatly,Wijewardane}:
\begin{equation}\label{ALDA+M}
V_{\rm xc}(z,t) = V_{\rm xc}^{\rm ALDA}(z,t)  + V_{\rm xc}^{\rm M}(z,t) \:,
\end{equation}
with the memory part
\begin{equation}
V_{\rm xc}^{\rm M}(z,t) = -\int_0^z \frac{dz'}{n(z',t)} \: \nabla_{z'} \sigma_{{\rm xc},zz'}(z',t) \:.
\end{equation}
The $zz$ component of the xc stress tensor is given by
\begin{equation} \label{timeintegral}
\sigma_{{\rm xc},zz'}(z',t)  = \int_0^t Y(n(z',t),t-t') \nabla_{z'} v_{z'}(z',t) dt' \:.
\end{equation}
Here, ${\bf v}(z,t) = {\bf j}(z,t)/n(z,t)$ is the time-dependent velocity field, where ${\bf j}(z,t)$ is the
current density. In 2D, the memory kernel $Y$ is given by
\begin{equation}
Y(n,t-t') =  \mu_{\rm xc} -\frac{n^2}{\pi} \int \frac{d\omega}{\omega} \: \Im f_{\rm xc}^L(\omega)
\cos[\omega(t-t')] \:,
\end{equation}
with the 2D xc shear modulus of the electron liquid \cite{QianVignale}
\begin{equation}
\mu_{\rm xc} = n^2 \left( \Re f_{\rm xc}^L(0) - (ne_{\rm xc})'' \right)
\end{equation}
(the prime denotes a derivative with respect to $n$).
In the following, we use the Holas-Singwi parametrization for the longitudinal frequency-dependent xc kernel
of the 2D electron liquid \cite{HolasSingwi}:
\begin{eqnarray} \label{Imfxc}
\Im f_{\rm xc}^L(\omega) &=& \frac{A \omega}{B^2 + \omega^2}\\
\Re f_{\rm xc}^L(\omega) &=& f_\infty  + \frac{A B}{B^2 + \omega^2} \:.
\end{eqnarray}
The coefficients $A$ and $B$ are given by
\begin{eqnarray}
A &=&  -\frac{11 \pi^2}{32} \\
B &=&  \frac{A}{(ne_{\rm xc})''-f_\infty} \:,
\end{eqnarray}
with
\begin{equation}
f_\infty = \frac{1}{2n} \left\{
-\frac{5}{2}\: n^2 \left( \frac{e_{\rm xc}}{n}\right)' + 12 n^{3/2} \: \left(
\frac{e_{\rm xc}}{\sqrt{n}}\right)' \right\}.
\end{equation}
It is easy to see that this simple parametrization for $f_{\rm xc}^L$ leads to zero shear modulus, $\mu_{\rm xc}=0$.
A more sophisticated interpolation formula, with finite $\mu_{\rm xc}$, has been derived by Qian and Vignale \cite{QianVignale},
but its input parameters are currently only available for a limited range of densities in the metallic regime. For our purposes,
it is therefore preferable to work with the Holas-Singwi formula (\ref{Imfxc}), which has the additional advantage that
it leads to a very simple expression for the memory kernel:
\begin{equation}
Y(n,t-t') = -\frac{An^2}{B} \: e^{-B(t-t')} \:,
\end{equation}
i.e., the system experiences an exponential memory loss. This is similar to what was observed \cite{UllrichTokatly,Wijewardane}
in 3D systems using the Gross-Kohn parametrization for $f_{\rm xc}^L$ \cite{GrossKohn}.
As a consequence, the numerical evaluation of the time integral in Eq. (\ref{timeintegral}) can be simplified
by introducing a cutoff in $t-t'$, i.e., not the entire history of the system from $t=0$ onwards needs to be included.

\section{Results and Discussion} \label{sec:results}

\subsection{Charge-density oscillations}

\begin{figure}
\unitlength1cm
\begin{picture}(5.0,8.25)
\put(-6.2,-9.2){\makebox(5.0,8.25){\includegraphics{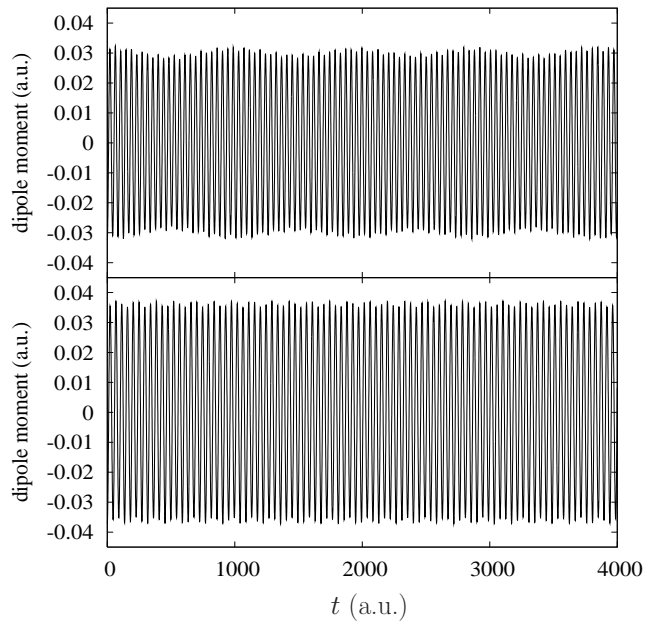}}}
\end{picture}
\caption{\label{figure3} Time-dependent dipole moment $d(t)$ associated with the charge-density oscillations in a
quantum strip of width $\Delta=10$, length $L=50$, and initial field strength $F=0.02$. Top: exact solution. Bottom: ALDA.
}
\end{figure}

\begin{figure}
\unitlength1cm
\begin{picture}(5.0,8.25)
\put(-6.2,-9.2){\makebox(5.0,8.25){\includegraphics{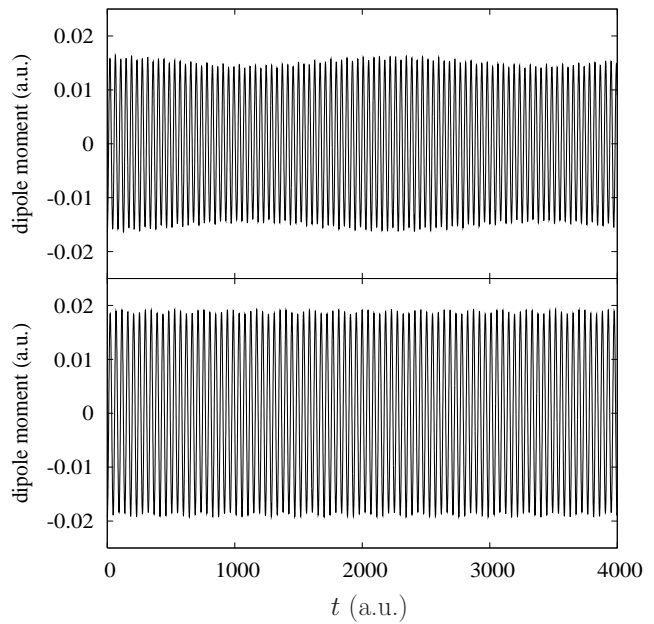}}}
\end{picture}
\caption{\label{figure4} Same as Fig. \ref{figure3}, but for a strip of length $L=100$.
}
\end{figure}

Figs. \ref{figure3} and \ref{figure4} compare the time-dependent dipole moment $d(t)$ [Eq. (\ref{TDdipole})] for quantum strips
of width $\Delta=10$ and lengths $L=50$ and $L=100$, respectively, calculated from the exact density (\ref{TDdensity}) and
from the ALDA density. The initial state was prepared with an external potential (\ref{external}) of  field strength
$F=0.02$, which was abruptly switched off at $t=0$.

At first sight, the ALDA charge-density oscillations seem to agree well with the exact ones, as far as the frequency
and the average amplitude of $d(t)$ are concerned. On closer examination, however, we observe
a beating pattern in the exact charge-density oscillations, which shows up as a low-frequency
modulation of the amplitude of $d(t)$. This effect is not reproduced by the ALDA, which produces a constant amplitude
(we disregard here the small, rapid wiggles in the amplitudes of $d(t)$, which are a nonlinear effect caused by
the relatively strong initial field, see Fig. \ref{figure2}).
In the following, we will focus on discussing the origin of these modulations,
and on the related shortcomings of the ALDA and the consequences thereof. It will turn out that this
effect provides a key to understanding the meaning of dissipation in TDKS theory.

In section \ref{timeexact} we considered the time evolution of the exact two-electron state, and
explained how it can be obtained by expanding the time-dependent wave function in the complete set
of field-free eigenstates, see Eq. (\ref{FieldFreeExp}). For the system parameters and field strengths under
consideration, this expansion turns out to be dominated by just a few leading terms in the summation over $A_j(t)\Phi_j^f$.
Looking at those few terms with the largest $|A_j(t)|^2$ will give us sufficient information to understand the electron dynamics
leading to the beating pattern in $d(t)$.

\begin{table}
\caption{\label{table1}
Leading terms in the expansion (\ref{FieldFreeExp}) of the time-dependent two-electron
wave function $\Phi(t)$ in terms of field-free eigenstates (for a strip with $\Delta=10$, $L=100$, and $F=0.02$).
$(\nu_1,\nu_2)$ indicates the dominating single-particle configurations (see Fig. \ref{figure5}).
}
\begin{ruledtabular}
\begin{tabular}{llcc}
$j$ & $(\nu_1,\nu_2)$&  $A_j^2$  & $E_j$ \\ \hline
1 &(1,1)       & $0.897406$ & $0$ \\
2 &(1,2),(2,1) & $0.098766$ & $0.148661$ \\
3 &(2,2)       & $0.002698$ & $0.294534$ \\
4 &(1,3),(3,1) & $0.001008$ & $0.394043$ \\
5 &(2,3),(3,2) & $0.000051$ & $0.542683$ \\
6 &(1,4),(4,1) & $0.000048$ & $0.739263$
\end{tabular}
\end{ruledtabular}
\end{table}

Let us analyze in detail the case $\Delta=10$, $L=100$ and $F=0.02$. We have solved the two-electron Schr\"odinger equation with
11 plane-wave and 8 standing-wave basis states, i.e.,  $|\kappa| \le 5$ and $\nu \le 8 $  in Eqs. (\ref{knum}) and  (\ref{nnum}).
Table \ref{table1} shows the six leading terms in the expansion (\ref{FieldFreeExp}) of
$\Phi(t)$ in terms of field-free eigenstates, i.e., those terms with the largest $A_j^2$, and the associated energies $E_j$, where
we define the field-free ground-state energy to be $E_1=0$. The remaining terms in Eq. (\ref{FieldFreeExp}) have values of $A_j^2$
that are orders of magnitude smaller.

\begin{figure}
\unitlength1cm
\begin{picture}(5.0,10.25)
\put(-6.2,-11.5){\makebox(5.0,10.25){\includegraphics{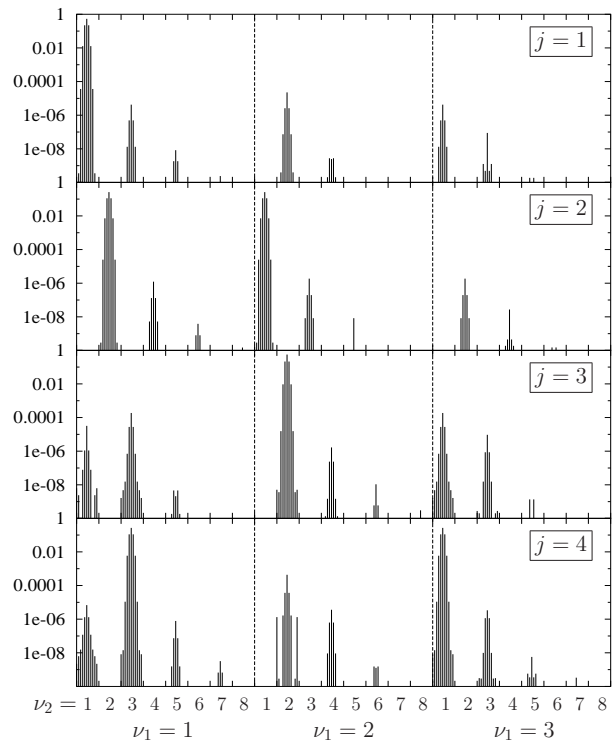}}}
\end{picture}
\caption{\label{figure5} Values of the coefficients $\left[C^{j,f}_{\nu_1\nu_2\kappa_1\kappa_1}\right]^2$
in the representation $\Phi_j^f= \sum_{\nu_1\nu_2 \atop \kappa_1\kappa_2} C^{j,f}_{\nu_1 \nu_2 \kappa_1 \kappa_2}
\Psi_{\nu_1 \nu_2 \kappa_1 \kappa_2}$ of the first four leading field-free eigenstates (see Table \ref{table1}).
Each configuration with standing waves $(\nu_1,\nu_2)$ along $z$ also has running waves along $x$ with  $(\kappa,-\kappa)$,
where $-5\le\kappa\le 5$, shown here as histograms. The dominating configurations are for $\kappa=0$, but finite
$\kappa$ are not negligible.
}
\end{figure}

According to equation (\ref{SlaterBasis}), each field-free eigenstate $\Phi_j^f$ is represented as a sum of single-particle
Slater determinants. The leading $\Phi_j^f$'s are dominated by configurations
$\Psi_{\nu_1\nu_2\kappa_1\kappa_2}$ whose standing-wave quantum numbers $(\nu_1,\nu_2)$ are given in
the second column of Table \ref{table1}, and which can have a broad range of plane waves $(\kappa_1,\kappa_2)$.
This is illustrated in detail in Fig. \ref{figure5}, which shows histograms of the coefficients
$\left[C^{j,f}_{\nu_1\nu_2\kappa_1\kappa_1}\right]^2$ in the expansion
$\Phi_j^f= \sum_{\nu_1\nu_2 \atop \kappa_1\kappa_2} C^{j,f}_{\nu_1 \nu_2 \kappa_1 \kappa_2}
\Psi_{\nu_1 \nu_2 \kappa_1 \kappa_2}$ of the first four leading field-free eigenstates.
One can clearly see that there are dominating pairs of standing-wave quantum numbers $(\nu_1,\nu_2)$, which explains
the assignment in
the second column of Table \ref{table1}. Each configuration with standing waves $(\nu_1,\nu_2)$ along $z$ is accompanied by left
and right running waves $(\kappa,-\kappa)$ along $x$. The case $\kappa=0$ is dominant, but finite $\kappa$ are not negligible.

To understand the beating pattern in the dipole oscillations of Figs. \ref{figure3} and \ref{figure4}, we now focus on the
first three leading field-free eigenstates and their standing-wave quantum numbers $(\nu_1,\nu_2)$, and for the moment disregard
the running waves along $x$. The beating pattern essentially arises from a superposition of two dipole oscillations
associated with the transitions $(1,1) \to (1,2),(2,1)$ and $(1,2),(2,1) \to (2,2)$. The associated energy differences are
$E_2 - E_1 = \omega_{21}=0.148661$ and $E_3 - E_2 = \omega_{32}=0.145873$. The two oscillation frequencies $\omega_{21}$
and $\omega_{32}$ are very close, and their difference $\omega_{21}-\omega_{32}=0.002788$ is precisely the frequency of the
amplitude modulation of $d(t)$. The resulting modulation period is $T_{\rm mod}= 2\pi/(\omega_{21}-\omega_{32}) = 2254$, which agrees
extremely well with the data shown in Fig. \ref{figure4}. Similarly, for the case $L=50$ shown in Fig. \ref{figure3}
we find $T_{\rm mod} = 964$ (here, the difference $\omega_{21}-\omega_{32}$ is a bit bigger).
The amplitude of the modulations of $d(t)$ depends on the field strength $F$ and remains small as long as
$A_1^2,A_2^2 \gg A_3^2$.

It is now easy to see why the ALDA misses the beating pattern in $d(t)$: the reason is that it does
not account for doubly-excited configurations. The ALDA includes only single excitations, which are the only
possible excitations of the KS system. Thus, transitions involving the $(\nu_1,\nu_2)=(2,2)$ configuration, which
are crucial to explaining the modulation of $d(t)$, cannot occur, and therefore no superposition effect takes place.

In addition to the standing-wave double excitations, the contribution of doubly excited running-wave states
$(\kappa,-\kappa)$ along $x$ are also important.
Again, the ALDA only includes the case $\kappa =0$ (single excitations along $x$ are not possible due to momentum
conservation). On the other hand, ignoring the states $(\kappa,-\kappa)$ with finite $\kappa$ in the expansion of the
full two-electron wave function would lead to substantially different energies $E_j$, and the low-frequency beating
pattern of $d(t)$ would be destroyed.

The exact xc potential (which we will construct in the next subsection) has to compensate for the absence of multiple
excitations in the TDKS wavefunction, and it does so through a nonadiabatic contribution. This is known from
linear response theory \cite{Maitra1}, where the xc kernel must have a frequency dependence to describe double
excitations.

\subsection{Exact xc potential and time-dependent energy}

\subsubsection{Construction of the exact time-dependent xc potential}

If the density $n({\bf r},t)$ of a system of two electrons in a singlet state is
given, it is a straightforward affair to construct that xc potential
$V_{\rm xc}({\bf r},t)$ which, when employed in a TDKS equation, reproduces this density \cite{DAmico}.
The doubly occupied TDKS orbital can be written as
\begin{equation} \label{phi2}
\varphi({\bf r},t) = \sqrt{ \frac{n({\bf r},t)}{2}} \: e^{i\alpha({\bf r},t)} \;,
\end{equation}
where the phase $\alpha$ is a real function and related to the current density as follows:
\begin{equation}
\nabla \alpha({\bf r},t) = {\bf j}({\bf r},t)/n({\bf r},t) \:.
\end{equation}
Inserting the ansatz (\ref{phi2}) into the TDKS equation, one obtains
\begin{equation}
V_{\rm xc}({\bf r},t) = V_{\rm xc}^{\rm stat}({\bf r},t) + V_{\rm xc}^{\rm dyn}({\bf r},t)\:.
\end{equation}
The first term,
\begin{eqnarray}
V_{\rm xc}^{\rm stat}({\bf r},t) &=& \frac{1}{4} \nabla^2 \ln n({\bf r},t) + \frac{1}{8}
|\nabla \ln n({\bf r},t)|^2 \nonumber\\
&-& V({\bf r},t) - V_{\rm H}({\bf r},t)\:,
\end{eqnarray}
is identical to the expression for constructing the static xc potential from a given static two-electron
density \cite{Filippi},
except that all quantities are now taken as time-dependent.
The second term,
\begin{equation}
V_{\rm xc}^{\rm dyn}({\bf r},t) = -\dot{\alpha}({\bf r},t) - \frac{1}{2} |\nabla \alpha({\bf r},t)|^2 \:,
\end{equation}
has no static counterpart and is therefore a truly dynamical contribution.

\subsubsection{The time-dependent energy}

Let us now consider the time dependence of the total energy $E(t)$ of the two-electron system. Since we are dealing with
free charge-density oscillations of a finite 2D quantum strip, i.e., there is no time-dependent external force,
the total energy of the full many-body system must obviously be constant.
This is easy to see for the exact time-dependent two-electron wave function $\Phi(t)$: according to equations
(\ref{FieldFreeExp}) and (\ref{AJ}), we simply have $E(t) = \sum_j E_j $ for all times.

On the other hand, it is less obvious what the TDDFT expression for the exact time-dependent total energy of
a many-body system should be (although some components of the TDDFT energy have been studied
in Ref. \cite{Hessler}).
However, for the purpose of this paper it is sufficient to define a quantity which we call the
{\em adiabatic energy}, $E_a(t)$. For a two-electron KS system with a doubly occupied single-particle
orbital, we have \cite{DAgosta}
\begin{eqnarray} \label{EKS}
E_a(t) &=& 2\int d{\bf r} \: \varphi^*({\bf r},t)\left[ \frac{1}{2}
\left( \frac{\nabla}{i} + {\bf A}_{\rm xc}({\bf r},t)\right)^2 \right.\nonumber\\
&+& V({\bf r},t) \bigg]\varphi({\bf r},t)+ E_{\rm H}[n(t)] +  E_{\rm xc}[n(t)] \;.
\end{eqnarray}
Here, $E_{\rm H}[n]$ is the Hartree energy functional, $E_{\rm xc}[n]$ is the ground-state xc energy functional
that was used
in the calculation of the initial state for the TDKS time propagation, and ${\bf A}_{\rm xc}$ is the nonadiabatic
xc vector potential. Clearly, $E_a(t)$ is not the true energy of the many-body system, but it reduces to the
ground-state energy in the static limit.
It can then be shown that the rate of change of the adiabatic energy is \cite{DAgosta}
\begin{equation}
\dot{E}_a(t) = \int d{\bf r} \: {\bf j}({\bf r},t) \cdot \dot{\bf A}_{\rm xc}({\bf r},t) \:.
\end{equation}
In the case of our 2D quantum strip, the density is spatially inhomogeneous along the $z$ direction
only, which means that we can replace the xc vector potential by the dynamical scalar potential, using
the relation
\begin{equation}
\dot{\bf A}_{\rm xc}({\bf r},t) = -\nabla V_{\rm xc}^{\rm dyn}({\bf r},t) \:.
\end{equation}
Furthermore, the $z$-components of the physical current $j_z(z,t)$ and the KS current
$j_{{\rm KS},z}(z,t)=2\Im(\varphi^* d\varphi/dz)$ of this particular two-electron system
become identical. Therefore, we obtain
\begin{equation} \label{EKSt}
\dot{E}_a(t) = -L \int_0^\Delta dz\: j_{z}(z,t)
\frac{d}{dz} \:V_{\rm xc}^{\rm dyn}(z,t) \:.
\end{equation}
This shows that the rate of change of the adiabatic energy is determined by the work done by the
forces associated with the dynamical xc potential.

In the case where we start from the exact time-dependent density and ask what the associated exact
adiabatic energy is, a direct evaluation of expression (\ref{EKS}) is not possible, since one doesn't know the
form of the exact xc energy functional. Fortunately, the adiabatic energy can easily be obtained from
Eq. (\ref{EKSt}) through a simple time integration:
$E_a(t) = \int_0^t \dot{E}_a(t') dt'$. This is a very convenient way of determining
the exact adiabatic energy (up to an irrelevant constant) from the exact density.

We have calculated $E_a(t)$ for the
exact solutions of the charge-density oscillations for the 2D quantum strips with $L=50$ and $L=100$,
see top parts of Figs. \ref{figure3} and \ref{figure4}. As expected, the exact adiabatic energy is {\em not}
constant, but rather rapidly fluctuating with time. For the sake of clarity, and since these rapid
fluctuations are not what we are primarily interested in, we define
a cycle-averaged adiabatic energy $\bar{E}_a(t)$, which is
obtained by averaging the adiabatic energy at each time $t$ over a
time window of one period of the charge-density oscillation (duration $\sim 40$ a.u.).

\begin{figure}
\unitlength1cm
\begin{picture}(5.0,8.25)
\put(-6.2,-9.2){\makebox(5.0,8.25){\includegraphics{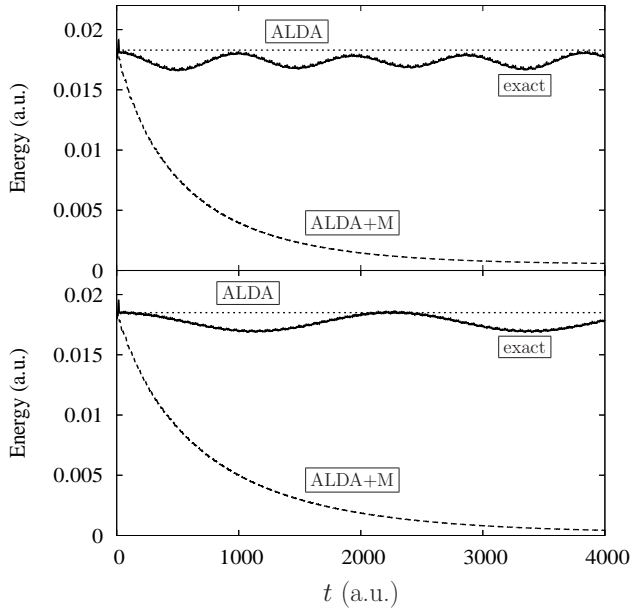}}}
\end{picture}
\caption{\label{figure6} Full lines: cycle-averaged adiabatic energy $\bar{E}_a(t)$ for the
exact solutions of the charge-density oscillations on the 2D quantum strip with $L=50$ (top) and $L=100$ (bottom).
Short and long dashed lines: $E_a(t)$ calculated with
ALDA and ALDA+M. For clarity, all energies are shifted so that they initially coincide.
}
\end{figure}

Figure \ref{figure6} shows the adiabatic energy for the two quantum strips, calculated with
different methods [both the ALDA and ALDA+M energies have been directly
determined from Eq. (\ref{EKS}); the ALDA+M results will be discussed in the next subsection].
One clearly sees slow oscillations of the exact $\bar{E}_a(t)$
with the same period as the amplitude modulations of $d(t)$
(see Figs. \ref{figure3} and \ref{figure4}). By contrast,
$E_a(t)$ in ALDA is constant as expected, since the ALDA dipole amplitudes are not modulated.

These results provide some interesting insights into TDDFT. It follows from the Runge-Gross
theorem \cite{rungegross} that there exists a unique TDKS system which reproduces any (reasonably well-behaved)
time-dependent density $n({\bf r},t)$ of a many-body system. In particular, if the exact functional for the xc potential
$V_{\rm xc}({\bf r},t)$ is used, the TDKS system gives the exact time-dependent density. However,
the adiabatic energy (as defined above) is not required to remain constant like
the true energy of the many-particle system (in the absence of an external field, of course).

How does this happen in our 2D model system? The exact TDKS system has to somehow reproduce the amplitude modulations
of the time-dependent dipole moment, but it cannot do so through a simple superposition of oscillations
associated with single and double excitations, as it happens in the full two-electron Schr\"odinger equation:
there are no double excitations in the TDKS system. Instead, the TDKS system has to produce the beating
pattern in $d(t)$ through the action of the xc potential. In other words, the dynamical part of the
xc potential, $V_{\rm xc}^{\rm dyn}$, acts in a sense like an ``external'' potential which periodically
drives and damps the charge-density oscillations of the system in order to increase or diminish the amplitude
of $d(t)$. From this point of view, $V_{\rm xc}^{\rm dyn}$ thus periodically dissipates
energy from the KS system, and then pumps it back into it. The adiabatic energy $E_a(t)$ serves as an indicator
for this behavior.

\subsection{Memory effects and  dissipation}

In the previous subsections, we have seen that the ALDA fails to reproduce some key features of the
dynamics of the two-electron system, related to the fact that it misses the double excitations.
We have also seen that the exact xc potential makes up for this through nonadiabatic contributions.
Let us now see how the nonadiabatic ALDA+M approximation of C-TDDFT performs, which was described in Section \ref{timeevol}.

\begin{figure}
\unitlength1cm
\begin{picture}(5.0,8.25)
\put(-6.2,-9.2){\makebox(5.0,8.25){\includegraphics{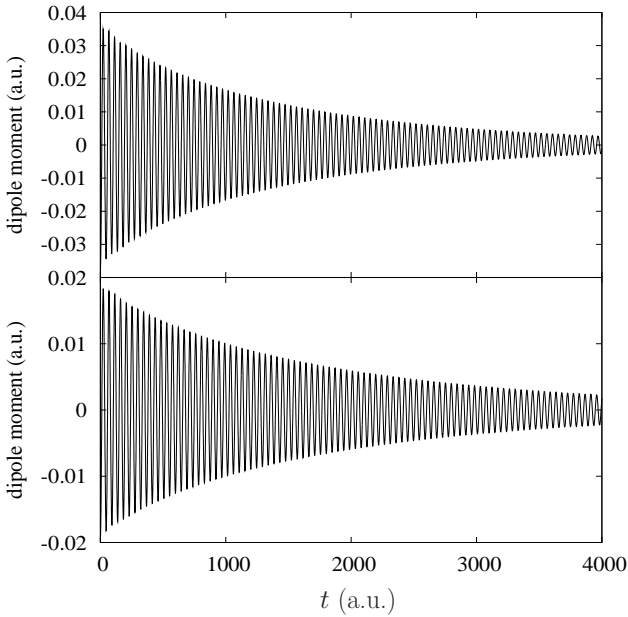}}}
\end{picture}
\caption{\label{figure7} Time-dependent dipole moment calculated with ALDA+M, for the same charge-density oscillations
treated in Figs. \ref{figure3} and \ref{figure4}. Upper panel: $L=50$, lower panel: $L=100$.
}
\end{figure}

Figure \ref{figure7} shows the time-dependent dipole moment for the charge-density oscillations of the same 2D
quantum strips discussed above, but now calculated using the nonadiabatic ALDA+M
xc potential [Eq. (\ref{ALDA+M})]. In both cases, the dipole oscillations are exponentially damped, similar to
what was previously observed for plasmon oscillations in a semiconductor quantum well \cite{Wijewardane}.
Representing the dipole moment as $d(t) \sim d_0 \cos(\omega t) e^{-\Gamma t}$, we find a
damping rate of about $\Gamma=0.0007$ for the $L=50$ strip and $\Gamma=0.00055$ for the $L=100$ strip.
As expected, the system with $L=100$ has a somewhat weaker damping, due
to its lower particle density, which reduces the probability of electron-electron scattering.

The long-dashed lines in Fig. \ref{figure6} show
the adiabatic  energy $E_a(t)$, calculated with ALDA+M. In both quantum strip systems, the energy is dissipated at an exponential rate,
$E_a(t) = E_a(0)e^{-2\Gamma t}$. As discussed in the previous subsection, the exact adiabiatic
energy oscillates, but on average there is no dissipation.
These results clearly show that the nonadiabatic ALDA+M functional fails for our finite two-electron system.
It results in an unphysical damping of the charge-density oscillations.

Technically, the dissipation
arises from the fact that $V_{\rm xc}^{\rm M}$ is a velocity-dependent potential. As was discussed in
Ref. \cite{UllrichTokatly}, the history-dependence of $V_{\rm xc}^{\rm M}$, which is governed by the
memory kernel $Y(n,t-t')$, is such that it accounts for both dissipative and elastic properties of
the electron liquid. The microscopic properties of the many-body system enter through the frequency-dependent
xc kernel of the homogeneous electron gas, $f_{\rm xc}(\omega)$. This function describes dynamical
processes of the homogeneous electron gas that go beyond single-particle excitations,
i.e., $f_{\rm xc}(\omega)$ contains the physics of multiple particle-hole excitations.
In principle, this should be a good thing, since we have seen from our model that the main
defect of the ALDA is the absence of double excitations. But in spite of this, the ALDA+M does not work
in our system. How can one understand this?

\subsection{Thermodynamic limit}

The central reason for the failure of the ALDA+M for finite systems is the fact that it is
based on the homogeneous electron gas, i.e., a reference system of infinite extent. Since it is a
local functional, the potential $V_{\rm xc}^{\rm M}(z,t)$ has exactly the same value at some position $x$
of our quantum strip as it would have if the strip were infinitely long, but with the {\em same} electron
density $n(z,t)$. Thus, by construction, ALDA+M treats all systems locally as if they were infinite, even if they're not.

We can carry out the thought experiment of increasing the length $L$ of the strip,
and simultaneously adding electrons to keep the same $n(z,t)$ as for the two-electron system. The ALDA
and the ALDA+M xc potentials would be unchanged, likewise the Hartree potential, and we would find exactly the
same charge-density oscillations across the strip. In particular, the damping in ALDA+M would be the same,
irrespective of the length of the strip.

On the other hand, the dynamics of the exact many-body system will change dramatically if we increase
both $L$ and the particle number. The time-dependent many-body wave function will contain not only single
and double excitations, but a vast number of multiple excitations. The density of levels in the excitation
spectrum will grow,
and eventually turn into a continuum. Recall that for the two-electron system, we explained the periodic
amplitude modulation of the charge-density oscillation through a superposition of two frequencies associated
with the dominant single and double excitation. If the system size grows, many more such
transitions will play a role, and we will have to form a coherent superposition of many close-lying oscillators.
The resulting beating pattern will become more complex, and seems difficult to predict.

However, we can get a clue from comparing the case of $L=50$ and $L=100$. For the longer strip, the
modulation period increases, i.e., the recurrence time becomes longer. In the limit
of infinite system size, this suggests that the recurrence time will in fact become infinite.
In other words, the charge-density oscillations will be irreversibly damped.

But where does the energy go?
In the free charge-density oscillations considered here, the wave function can be expressed as a linear
superposition of field-free many-body states, each of which carrying
a time-dependent phase $\exp(-i E_j^f t)$ [this is a generalization of Eq. (\ref{FieldFreeExp}) to $N$ particles]. Thus,
the energy is, from the very beginning when the oscillation is triggered, shared in a fixed manner among all
excited-state configurations that make up the time-dependent many-body wave function. In turn, the charge-density
oscillation is a coherent superposition of a continuum of single and multiple particle-hole excitations, which
steadily run out of phase. This reduces the amplitude of the collective mode due to destructive interference.

To give a simple illustration, consider
the case of an exponential damping, $d(t) = d_0  \cos(\omega t) e^{-\Gamma t}$. We can carry out a
Fourier analysis of the spectral content of $d(t)$, and the result is that $d(t)$ arises from
a superposition of a continuum of oscillators whose frequency distribution has a Lorentzian shape of half-width $\Gamma$
centered around $\omega$.

The ALDA+M functional of C-TDDFT, per construction, becomes exact in the limit of an extended system
whose ground-state density, as well as the inhomogeneity of the time-dependent perturbation, are slowly
varying in space \cite{VK,VUC,UllrichVignale}. For such a system, the damping of the plasmon amplitude will be
correctly described. In the many-body system, the damping occurs through interference within
the continuum of multiple excitations, but in C-TDDFT, it has to happen in a completely different way:
the nonadiabatic piece of the xc potential acts like an external damping force. The outcome, i.e.
the behavior of the time-dependent density  $n(z,t)$, is the same.

\section{Conclusion} \label{sec:conclusions}

We have presented a simple two-electron system which has the appealing feature of being exactly solvable
with modest computational cost, and rich enough to provide new insight into the problem of dissipation of collective
electron dynamics. From the point of view of the exact time-dependent many-body wave function, plasmon
dissipation occurs through a superposition of a continuum of oscillators, associated with transitions
between multiply excited states, which slowly and irreversibly run out of phase. The phenomenon can be viewed like
a beat, but with an infinitely long recurrence time. Consequently, there is no loss of energy in
the many-body system, and, in a sense, not even a re-distribution into other degrees of freedom.

From a TDDFT point of view, all we can say is that we have a time-dependent density which produces
a time-dependent dipole moment whose amplitude steadily decreases. The exact TDKS system accomplishes
this through a nonadiabatic xc potential which acts like a damping force. As a result, energy of the KS
system is lost, but this is the price we have to pay to reproduce the exact density.

The ALDA+M xc functional of C-TDDFT has been constructed for infinite systems, and becomes exact
in the appropriate limits \cite{VK}. For finite systems, on the other hand, it introduces a spurious damping
of electron dynamics. For example, if the method is applied to atoms, one obtains excitation energies
with finite linewidths \cite{UllrichBurke}. On the other hand, the static limit of the Vignale-Kohn
functional \cite{VK} seems to work well for polarizabilities of polymers \cite{Faassen}.
Thus, more tests of C-TDDFT are needed to explore its usefulness for practical applications.

However, it seems unlikely that a time-dependent xc functional based on the homogeneous electron gas
can correctly describe the subtle aspects of the dynamics of {\em both} finite and extended systems that we
have discussed in this paper. A more promising approach may be through orbital-based functionals such
as the time-dependent optimized effective potential \cite{TDOEP}, which will be the subject of
future studies.

\acknowledgments
This work was supported by NSF Grant No. DMR-0553485 and by Research Corporation.
Discussions with E. K. U. Gross, Robert van Leeuwen, Neepa Maitra, and Giovanni Vignale
are gratefully acknowledged.


\end{document}